\documentclass[reprint, amsmath,amssymb, aps, pra, longbibliography]{revtex4-1}

\usepackage{soul}
\usepackage{amsmath}   
\usepackage{amssymb}
\usepackage{graphicx}
\usepackage{dcolumn}
\usepackage{bm}
\usepackage{amsfonts}
\usepackage{subfigure}
\usepackage{array}
\usepackage{float}
\usepackage{color}
\usepackage[colorlinks=true,linkcolor=blue]{hyperref}%
\hypersetup{allcolors=blue}
\usepackage[normalem]{ulem}
\usepackage{xcolor}

\newcommand{\afix}[1]{\textcolor{red}{#1}}

\begin{document} 

\title{Nonadiabatic decay of metastable states on coupled linear potentials}
\date{\today }

\author{Alisher Duspayev}
    \email{alisherd@umich.edu}
\author{Ansh Shah\footnotemark[1]}  
    \thanks{A. D. and A. S.  contributed equally to this work}
\author{Georg Raithel}
\affiliation{Department of Physics, University of Michigan, Ann Arbor, MI 48109, USA}

\begin{abstract}
Avoided crossings of level pairs with opposite slopes can form potential energy curves for the external degree of freedom of quantum particles. 
We investigate nonadiabatic decay of metastable states on such avoided crossings (MSACs) using diabatic and adiabatic representations. The system is described by a single scaled adiabaticity parameter, $V$. The time-independent two-component Schr\"odinger equation is solved in both representations, and the nonadiabatic lifetimes of MSACs are determined from a wave-function flux calculation and from the Breit-Wigner formula, leading to four lifetime values for each MSAC. We also solve the time-dependent Schr\"odinger equation in both pictures and derive the MSAC lifetimes from wave-function decay. The sets of six non-perturbative values for the MSAC lifetimes agree well, validating the approaches. As the adiabaticity parameter $V$ is increased by about a factor of ten, the MSAC character transitions from marginally to highly stable, with the lifetimes increasing by about ten orders of magnitude. 
The $\nu$-dependence of the lifetimes in several regimes is discussed.
Time-dependent perturbation theory is found to yield approximate lifetimes that deviate by $\lesssim 30\%$ 
from the non-perturbative results, while predictions based on the semi-classical Landau-Zener tunneling equation are found to be up to a factor of twenty off, over the ranges of $V$ and $\nu$ studied. The results are relevant to numerous atomic and molecular systems with quantum states on intersecting, coupled potential energy curves.
\end{abstract}

\maketitle

\section{Introduction}
\label{sec:intro}
Potential wells emerging from two intersecting diabatic potentials with opposite slopes, coupled by an (approximately) constant interaction, are abound in physics and chemistry~\cite{nakamurabook, tully2004}. Examples include atom traps in optical lattices with Raman couplings~\cite{zhang2005, lundblad, pan2016, anderson2021}, confinement of Bose-Einstein condensates on RF-dressed magnetic potentials with spin-dependent slopes~\cite{leanhardt2002, zobay2004, white2006, hofferbeth2007}, atom interferometry in RF-dressed magnetic guiding potentials~\cite{hanselpra, sherlock, Vangeleyn_2014, Navez_2016}, dressed atom-RF-field states in cavity-QED systems~\cite{raithel1995, harochermp}, Rydberg atoms in external fields~\cite{rubbmark, gall, reinhard2007}, intersecting potential energy curves with radially dependent adiabatic electronic states in Rydberg-Rydberg~\cite{shafferreview, hollerith}, Rydberg-ground~\cite{shafferreview, feyreview} and Rydberg-ion~\cite{duspayev2021, deiss2021, zuber2021, duspayev2021nad} molecules, and a host of conical intersections in quantum chemistry~\cite{yarkony, matsika, domcke2011conical, malhado}. 
If the slopes of the diabatic potentials have opposite signs, the upper adiabatic potential surface exhibits a potential well, 
and the classical motion in this well is a bound, periodic oscillation about the avoided crossing. Such cases are common in molecular physics, as in Rydberg-ion molecules~\cite{duspayev2021, deiss2021, zuber2021, duspayev2021nad}, and in atom trapping~\cite{Colombe_2004, Garraway_2016, Burrows_2017}. The semi-classical Landau-Zener (LZ) tunneling equation~\cite{LZener, Zener}
has sometimes been applied to estimate nonadiabatic decay rates of quantum states in such adiabatic-potential wells. The LZ estimates are exponentially dependent on several parameters, including a fixed, classical mass-point velocity that is assumed to approximate the vibrational quantum motion. LZ estimates of nonadiabatic decay rates of quantum states with low vibrational quantum numbers
can differ significantly from their true quantum-mechanical values~\cite{Burrows_2017,duspayev2021nad}. 

In this paper, we present a non-perturbative, 
quantum-mechanical analysis of nonadiabatic decay of low-lying metastable states at avoided crossings (MSACs).
Similar descriptions have previously been employed to model wave-packet dynamics on intersections~\cite{arnold, Burrows_2017} and in Rydberg-ground molecules~\cite{hummelpra2021, hummelprl2021}. 
Here, we concentrate on the nonadiabatic lifetimes of quasi-stationary MSACs, which are 
important in the aforementioned applications. 
After explaining our model and the utilized techniques in Sec.~\ref{sec:Methods}, in Sec.~\ref{sec:res} we obtain solutions 
of the time-dependent and time-independent Schr\"odinger equations in 
both diabatic and adiabatic representations. 
We extract nonadiabatic MSAC lifetimes from six non-perturbative methods, and compare and interpret the results. 
The analysis is performed for a range of 
coupling strengths between the diabatic potentials, for MSACs with vibrational quantum numbers ranging up to about 15.
In Sec.~\ref{sec:approx}, we compare the non-perturbative MSAC lifetime results 
with estimates based on time-dependent 
perturbation theory, and with semi-classical estimates based on the LZ formula. The paper is concluded in Sec.~\ref{sec:concl}.

\section{Methods}
\label{sec:Methods}

\subsection{System under study}
\label{subsec:system}

In the system of interest, the physical Hamiltonian in the diabatic representation,
\begin{equation}
\hat{H}_p =  -\frac{\hbar^2}{2M}
\begin{pmatrix}
 \frac{d^2}{d x_p^2}& 0 \\
0 & \frac{d^2}{d x_p^2}
\end{pmatrix}
+ 
\begin{pmatrix}
-\frac{\alpha_p}{2} x_p & V_p \\
V_p     & \frac{\alpha_p}{2} x_p
\end{pmatrix} \quad  ,
\label{eq:h0}
\end{equation}
acts on a two-component wave function $( \psi_1(x), \psi_2(x) )$ with a position-independent, internal state space denoted as $\{ \vert 1 \rangle , \vert 2 \rangle\}$, in that order. The constants $\alpha_p$ and $V_p$ are chosen  positive 
and real. 
The effective particle mass is denoted $M$, and the external degree of freedom has a spatial coordinate $x_p$. The diabatic energies of the internal states $\{ \vert 1 \rangle, \vert 2 \rangle\}$ as a function of $x_p$ are $V_{1} = - \alpha_p x_p/2$ and $V_{2} = \alpha_p x_p/2$, respectively, with differential slope $\alpha_p$, and the constant coupling between these states is $V_p$.

For convenient description of different physical systems, we use the following units for length, energy, time and frequency,
\begin{eqnarray}
{\rm{length}} \, & : & \, l_0= \sqrt[\leftroot{-0}\uproot{6} \scriptstyle 3]{\frac{\hbar^2}{M \alpha_p}} \nonumber \\
{\rm{energy}} \, & : & \, w_0 = \frac{\hbar^2}{M l_0^2} \nonumber \\
{\rm{time}} \, & : & \, t_0 = \frac{\hbar}{w_0} \nonumber\\
{\rm{frequency}} \, & : & \, f_0 = \frac{w_0}{\hbar}
\label{eq:units}
\end{eqnarray}
Expressing length and energy in these units, the Hamiltonian in Eq.~\ref{eq:h0} transforms into the scaled Hamiltonian in diabatic representation,
\begin{equation}
\hat{H}_D =  -\frac{1}{2}
\begin{pmatrix}
 \frac{d^2}{d x^2}& 0 \\
0 & \frac{d^2}{d x^2}
\end{pmatrix}
+ 
\begin{pmatrix}
-\frac{1}{2} x & V \\
V     & \frac{1}{2} x
\end{pmatrix} \quad  ,
\label{eq:hs}
\end{equation}
with scaled position $x=x_p/l_0$ and scaled coupling strength
\begin{equation}
V = \frac{V_p}{w_0}
\label{eq:coupls}
\end{equation}
The characteristic half 
width of the crossing region in scaled units
is $x_w = 2 V$; in physical length units it is $x_{wp} = V l_0 = 2 V_p/\alpha_p$. The scaled coupling $V$ serves as an adiabaticity parameter: the larger $V$, the more adiabatic a system will behave, and the less affected the MSACs will be by nonadiabatic decay. In the following, we will use the scaled units defined in Eq.~\ref{eq:units}. 

\begin{figure*}[thb]
 \centering
  \includegraphics[width = 0.98\textwidth]{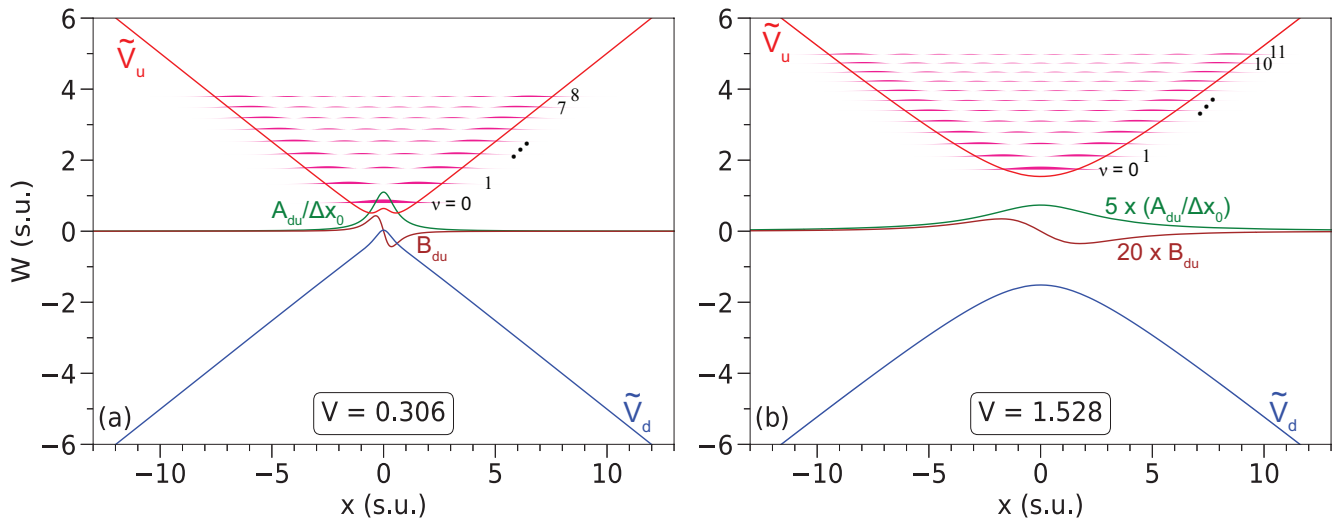}
  \caption{(Color online) PECs and nonadiabatic couplings in the Hamiltonian in Eq.~(\ref{eq:ha}) for the cases $V = 0.306$ (a) and $V = 1.528$ (b). In order to show  the first-order coupling, $A_{du}$, on a physically relevant energy scale, we plot $A_{du}$ divided by the position uncertainty of the ground state in harmonic approximation, $\Delta x_0 = V^{1/4}$. Note the scaling factors for the nonadiabatic couplings in (b). 
  The wave-function densities of the lowest 9 (a) and 12 (b) MSACs are also shown. The baselines of the individual wave function plots correspond with the respective resonance energies, $W_\nu$, on the vertical axis.}
  \label{figure1}
\end{figure*}

The $x$-dependent adiabatic-state basis $\{ \vert u \rangle , \vert d \rangle \}$ for the internal degree of freedom, and the adiabatic potentials $V_u$ and $V_d$, are defined by $\hat{H}_D \vert u \rangle = V_u (x) \vert u \rangle$ and $\hat{H}_D \vert d \rangle = V_d (x) \vert d \rangle$, with $u$ and $d$ standing for ``up'' and ``down'' in energy, $V_u$ positive, and $V_d = -V_u$. With the notation $\vert u \rangle (x) 
= \sum_{i=1,2} \chi_{u,i} (x) \vert i \rangle$ and $\vert d \rangle (x)
= \sum_{i=1,2} \chi_{d,i} (x) \vert i \rangle$, the first- and second-order nonadiabatic couplings are 
\begin{eqnarray}
A_{\alpha, \beta}(x) & = & - \sum_{i=1,2} \chi^*_{\alpha, i} (x)  \frac{d}{d x} \chi_{\beta, i}(x) \nonumber \\
B_{\alpha, \beta}(x) & = & - \frac{1}{2} \sum_{i=1,2} \chi^*_{\alpha, i} (x)  \frac{d^2}{d x^2} \chi_{\beta, i} (x) \quad,
\end{eqnarray}
There, the index $i$ denotes diabatic and the Greek letters adiabatic basis states. The $\chi_{\alpha, i}$ can be chosen real. The 
2x2 matrix $A_{\alpha, \beta}$ then is anti-symmetric at any value of $x$, with $\alpha$ and $\beta$ being $u$ or $d$. The diagonal elements of $B_{\alpha, \beta}(x)$ are compounded with the adiabatic potentials to yield the potential energy curves (PECs) $\tilde{V}_\alpha(x) = V_\alpha (x) + B_{\alpha, \alpha} (x)$, with $\alpha=u$ or $d$. The adiabatic Hamiltonian, which is a special case of the Born-Huang representation~\cite{BHAbook, Agostinireview} for the case studied in our paper, then writes
\begin{eqnarray}
\hat{H}_A & = & -\frac{1}{2}
\begin{pmatrix}
 \frac{d^2}{d x^2}& 0 \\
0 & \frac{d^2}{d x^2}
\end{pmatrix}
+ 
\begin{pmatrix}
\tilde{V_u}(x) &  0 \\
0    & \tilde{V_d}(x) 
\end{pmatrix} 
\nonumber \\
~ & ~ & +
\begin{pmatrix}
0  & B_{ud}(x) + A_{ud}(x) \frac{d}{d x} \\
     B_{du}(x) - A_{ud}(x) \frac{d}{d x} & 0 
\end{pmatrix} 
\quad  .
\label{eq:ha}
\end{eqnarray}  
This Hamiltonian acts on the adiabatic wave functions, $(\psi_u (x), \psi_d (x))$. As visualized in Fig.~\ref{figure1}, the nonadiabatic $A$- and $B$- couplings vanish for $x \gg V$, with $V$ from Eqs.~\ref{eq:hs} and~\ref{eq:coupls}.

\subsection{Time-independent solutions}
\label{subsec:tindep}

\subsubsection{Diabatic representation}
\label{subsubsec:tindepdia}

A straightforward method to arrive at a non-perturbative solution is to solve the time-independent Schr\"{o}dinger equation (TIDSE) for the
Hamiltonian from Eq.~\ref{eq:hs}. Here, we are interested in the energy range $W>V$, where  MSAC resonances exist. The MSAC resonance energies and corresponding two-component wave functions are obtained numerically. As a spatial integration method, we have chosen a 4-th order Runge-Kutta (RK) method, which allows for first-derivative terms (needed  in the adiabatic representation discussed in Sec.~\ref{subsubsec:tindepadia}).  In the following, relevant details are explained. 

It is well-known from textbooks that for a spin-less particle on a linear potential the wave-function solutions are given by Airy functions (see, e.g.,~\cite{griffithsqm}). In the case of two coupled linear potentials, as in Eq.~\ref{eq:hs}, the matching of the boundary conditions in the classically forbidden regions turns out to be numerically delicate due to the coupling $V$ between the classically-allowed, Airy-function-like
solutions to the co-located classically forbidden ones. In the asymptotic regions, the allowed solutions are, locally, approximately given by $a(x) \cos(k x + \phi)$, with a slowly-varying local amplitude $a(x)$, wave number $k(x)=\sqrt{2(|x|/2 + W)}$, the quantum state's scaled energy $W$, and a  phase $\phi$. For large $|x|$, the classically-forbidden solutions are then approximately given by $-\frac{a(x) V}{|x|/2 +W} \cos(k x + \phi)$. The amplitudes of the forbidden solutions drop off quite slowly in $|x|$, because $V$ is fixed and never ``turns off''. In the numerical implementation, this exacerbates the tendency of the classically-forbidden solutions to exponentially diverge at large $\vert x \vert$. The issue is addressed by choosing sufficiently small values for the spatial step size, $\Delta x$, and for the slope iteration parameter, $s$, explained in the next paragraph. The issue is less pronounced in the adiabatic approach, because the nonadiabatic $A$- and $B$-couplings both do ``turn off'' at large  $\vert x \vert$ (see Sec.~\ref{subsubsec:tindepadia}).

The energy spectrum of the Hamiltonian in 
Eq.~\ref{eq:hs} is continuous and ranges from
$-\infty$ to $\infty$. The numerical treatment is simplified by the symmetry of the real-valued solutions. For each energy $W$ there exists an even and an odd solution. Even solutions, which are associated with even-parity MSACs, are of the form $\psi_1(x) = 1 - s x$ and  $\psi_2(x) = 1 + s x$ for $\vert x \vert \rightarrow 0$, with a slope parameter  $s$.  The odd solutions are of the form  $\psi_1(x) = 1 + s x$ and  $\psi_2(x) = - 1 + s x$ for $\vert x \vert \rightarrow 0$. Further, for any $x$ it is $\psi_2(-x) = \psi_1(x)$ for the even and $\psi_2(-x) = - \psi_1(x)$ for the odd solutions. For any  
energy $W$, this leaves only one parameter - the slope $s$ - to be iterated.
In both even and odd cases, the slope parameter $s$ is iterated to minimize the classically
forbidden wave-function components at the chosen spatial-range limit, $|x| = x_{max}$.
We vary $x_{max}$ depending on $V$ and $W$, so as to allow for maximum outward propagation before the wave functions diverge due to numerical inaccuracies. For each energy $W$, this procedure yields exactly one even and one odd solution. 

\subsubsection{Adiabatic representation}
\label{subsubsec:tindepadia}

TIDSE in the adiabatic picture has the Hamiltonian from Eq.~\ref{eq:ha}. The even adiabatic solutions are of the form $\psi_{u}(x) = 1 $ and $\psi_{d}(x) = s x $ for $|x| \rightarrow 0$, and the odd ones are of the form $\psi_{u}(x) = s x $ and $\psi_{d}(x) = 1 $ for $|x| \rightarrow 0$.
As in Sec.~\ref{subsubsec:tindepdia}, the slope parameter $s$ is iterated to minimize the classically
forbidden wave-function components $\psi_u(x)$ at the spatial-range limits, $|x| = x_{max}$. For each energy value $W$, there exist exactly one even and one odd solution. In the numerical treatment, the tendency of the classically forbidden solutions on the respective PECs to exponentially diverge at large $\vert x \vert$ is less pronounced in the adiabatic representation than it is in the diabatic representation (Sec.~\ref{subsubsec:tindepdia}), because the nonadiabatic $A$- and $B$-couplings ``turn off'' at large  $\vert x \vert$. In contrast, in the diabatic representation the constant coupling $V$ does not ``turn off'' at large $|x|$.

\begin{figure*}[thb]
 \centering
  \includegraphics[width=0.98\textwidth]{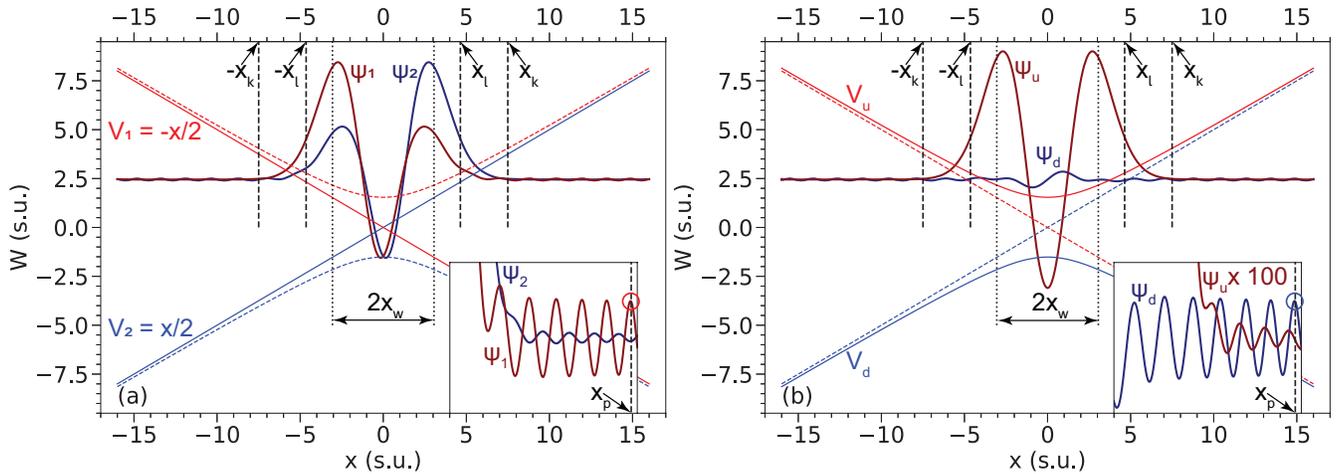}
  \caption{(Color online) Details for the lifetime calculations 
  in diabatic (a) and adiabatic (b) representations, with diabatic potentials $V_{1/2} = \mp x/2$ and adiabatic potentials $V_{u}$ and $V_d$ [all quantities
  in scaled units (s.u.)]. The displayed case is for a coupling strength of $V=1.5275$~s.u.. The wave functions $\psi_1$ and $\psi_2$ in (a), and $\psi_u$ and $\psi_d$ in (b), show the vibrational MSAC $\nu = 2$. The insets in (a) and (b) show magnified views of the wave-function tails in the classically forbidden regions of the respective higher-energy potentials. Note that $\psi_{u}$ in the inset in (b) is multiplied by a factor of 100. The markers $x_l$, $x_k$ and $x_w$ in the main plots as well as $x_p$ and the open circles in the insets are to illustrate details of the flux calculations explained in the text.}
  \label{figure2}
\end{figure*}

\subsubsection{MSAC resonances}

\label{subsubsec:resonances}

The energies $W_{\nu}$ of the MSAC resonances, labeled by an integer vibrational quantum number $\nu$,  
can be determined iteratively by locating the 
energy values at which the amplitudes of the sinusoidal wave-function tails in the respective classically-forbidden regions become minimal near the edges of the spatial integration range,  $|x| = x_{max}$. 
We label the resonances starting with $\nu=0$ for the MSAC ground state. 
The coupling parameter $V$ is varied between 0.3 (least adiabatic) and 2.8 (most adiabatic), in scaled units as defined in Eq.~\ref{eq:units}. 
We find all MSAC resonances within an energy range of about $V < W \lesssim V + 3.8$. For $V$ ranging between 0.3 and 2.8, the number of MSACs with $V < W_{\nu} \lesssim V + 3.8$  ranges from $\nu+1=9$ to 15, respectively. The integration limit, $x_{max}$, is shifted outward with increasing $V$ and $\nu$ in order to locate the MSAC energies as accurately as possible over the entire $V$- and $\nu$-range studied. Here, $x_{max}$ is varied between $x_{max}=13$ at the lowest $V$ and $\nu$, and $x_{max}=19$ at the largest $V$ and $\nu$. 

For illustration, in Fig.~\ref{figure1} we show plots of the adiabatic potentials $\tilde{V}_d$ and $\tilde{V}_u$ and the $A$- and $B$-potentials for $V=0.306$ and $V=1.5275$, as well as the obtained lowest MSACs. In addition to $A_{ud}(x) = -A_{du}(x)$, in the present problem it is also $B_{ud}(x) = -B_{du}(x)$. Fig.~\ref{figure1} illustrates 
the rapid drop in amplitude of both the $A$- and $B$-potentials with increasing $V$. The effect of the diagonal $B$-potentials, $B_{dd}(x)$
and $B_{uu}(x)$, only becomes apparent in the $V=0.306$-case in the form of small humps in the range $|x| \lesssim 1$.
The $A$-couplings generally appear to be more important than the $B$-couplings, as confirmed directly in Sec.~\ref{subsec:FGR}.  
A feature that becomes important in the interpretation of the $\nu$-dependence of the MSAC lifetimes in 
Sec.~\ref{subsec:rsurvey}
is that at low $V$ the approximate reach of the $A$- and $B$-potentials, given by the crossing half width, $x_w = 2 V$,  is smaller than
the typical wave function extents, whereas at large $V$  the $A$- and $B$-potentials are spread out over 
the entire typical wave-function extent.

\subsubsection{MSAC lifetimes from flux calculation}
\label{subsubsec:flux}

The main interest in the present work is in the nonadiabatic lifetime of the lowest MSAC resonances. 
To that end, we compute even and odd solutions on a dense grid of the continuous energy $W$, 
and determine the resonance centers, $W_\nu$, as described in Sec.~\ref{subsubsec:resonances}. 
In either representation, the MSAC resonances correspond with wave-function solutions 
that minimize the amplitudes of the sinusoidal wave-function tails in the respective classically forbidden regions at the edges of the spatial integration range, $\pm x_{max}$. As seen in Fig.~\ref{figure2}, in the asymptotic regions both classically allowed and forbidden wave-function tails are locally of the form $\psi(x) = a(x) \cos[k(x) x + \phi(W)]$, with a slowly-varying amplitude $a$, an energy-dependent phase $\phi$, and a slowly-varying wave number $k$. In the diabatic representation, the oscillatory behavior of the classically-forbidden tails results from the fixed coupling $V$, which induces $\pi$-out-of-phase classically-forbidden tails. Denoting the amplitude of the classically allowed tail $a_a$, and that of the co-located classically forbidden tail $a_f$, at the spatial integration boundary $x_{max}$ the amplitude $a_f \sim \frac{a_a V}{x_{max}/2 +W}$ can be on the order of 10$\%$ of $a_a$ [see, for instance, the inset of Fig.~\ref{figure2}~(a)]. In the adiabatic representation, the oscillatory behavior of the classically-forbidden tails primarily results from the diminishing nonadiabatic $A$-coupling, which induces $\pi/2$-out-of-phase forbidden tails with much smaller amplitudes than in the diabatic representation [note that in the inset in Fig.~\ref{figure2}~(b) the classically forbidden tail of $\psi_u$ is magnified by a factor of 100]. 

The steady-state solutions can be viewed as superpositions of in-going ``pump'' waves with out-going back-scattered waves,
forming perfect standing waves on either side of the potential. 
For a scalar wave function with oscillatory tails of amplitude $a$ in the positive- and negative-$x$ domains, $\psi(x) =
(a/2) [\exp( {\rm{i}} (kx + \phi)) + \exp( - {\rm {i}} (kx + \phi)]$, the outgoing 
flux, $j= \frac{1}{2i} (\psi^* \frac{d}{d x}\psi - \psi \frac{d}{d x}\psi^*)$, summed over the positive- and negative-$x$ domains, is $2 k |a|^2$. 
If the wave function contains a metastable resonance in a central potential well, 
the decay rate of the resonance upon turning off the pump waves is $\Gamma = 2 k |a|^2 / P_0$, where $P_0$ is the wave-function norm. 
Since $P_0$ is proportional to $|a|^2$, the factor $|a|^2$ drops out.

We first discuss the implementation in diabatic representation, in which the norm $P_0$ of the MSAC wave functions $P_0 = \int_{-x_k}^{x_k}
[|\psi_1(x)|^2 + |\psi_2(x)|^2] dx$. We define the integration boundary $x_k$ via $\int_{x_l}^{x_k} k(x) dx = r_k$, where $x_l$ is the positive classical turning point of the state of interest on $V_u(x)$, $x_k > x_l$, and $k(x) = \sqrt{2 (V_u(x) - W)}$. 
In this way, the limit $x_k$ is set such that $P_0$ captures the decaying tails of the MSAC resonances in the classically-forbidden regions 
to within $r_k$ semi-classical $1/e$ decay lengths outside the classical turning points. Here we use $r_k=3$, which is large enough for $P_0$ to capture the entire resonance norm, and small enough to not include substantial probability from the oscillatory wave-function tails. 
The exact value of $r_k$ is not important. Note that we define the boundaries via the upper {\sl{adiabatic}} potential, $V_u(x)$, in {\sl{both}} the diabatic and adiabatic representations. The limits $x_l$ and $x_k$ are visualized in Fig.~\ref{figure2}.

The amplitudes $a_1$ and $a_2$ of the oscillatory wave-function tails are found by first locating a position $x_p$ 
where the (classically allowed) tail of $\psi_1(x)$ takes an extremal value close to the positive limit of the spatial integration range, $x_{max}$ [see circle in the inset of Fig.~\ref{figure2}~(a)]. Near $x_p$, both wave functions $\psi_i(x)$ then are of the form $\psi_i(x) = a_i \cos[k_i (x_p) x + \phi_i]$, with $i=1 , \, 2$. Using three adjacent carrier points of each $\psi_i(x)$ with $x_p$ at the center, 
we compute the wave numbers
$k_i (x_p) = \sqrt{ \vert \frac{\frac{d^2} {dx^2} \psi_i(x_p) }{\psi_i(x_p)} \vert }$
and the amplitudes $a_i (x_p) = \sqrt{\psi_i(x_p)^2 + ( \frac{d}{dx}\psi_i (x_p) / k_i(x_p))^2}$.
Due to the position-independence of the diabatic internal states, $\{ \vert 1 \rangle , \vert 2 \rangle\}$, the fluxes in the two wave-function components just add up. The $1/e$ lifetime $\tau_{nad, FC}=1/\Gamma_{nad, FC}$, obtained from the time-independent MSAC wave function in the diabatic picture, then is given by

\begin{equation}
\tau_{nad, FC} = \frac{1}{\Gamma_{nad, FC}}  =  \frac{P_0}{2 (k_1 a_1^2 + k_2 a_2^2)}
 \quad .
\end{equation}
It is noted that $k_1 \approx k_2$ in all cases, whereas the ratio $a_2/a_1$ increases with $V$. 

We also obtain the lifetimes in the adiabatic representation, in which the wave-function computation is numerically more stable. 
The norm integral $P_0$ is computed with the same boundary $x_k$ as in the diabatic representation, 
$P_0 = \int_{-x_k}^{x_k} [|\psi_u(x)|^2 + |\psi_d(x)|^2] dx$. We first find a peak location $x_p$ of the (classically-allowed) tail of $\psi_d(x)$ near the integration limit, $x_{max}$ [see circle in the inset of Fig.~\ref{figure2}~(b)]. In the flux calculation in the adiabatic 
representation, the $x$-dependence of the adiabatic internal-state basis $\{ \vert u \rangle (x), \vert d \rangle (x) \}$
must be considered. Therefore, the adiabatic two-component wave function $(\psi_d(x), \psi_u(x))$ 
is transformed into diabatic representation, $(\psi_1(x), \psi_2(x))$ at three adjacent $x$-values centered at $x_p$. 
The decay rate $\Gamma_{ad, FC}$ and the $1/e$ lifetime $\tau_{ad, FC}$ of the time-independent MSAC wave function in the adiabatic picture are then computed from $(\psi_1(x), \psi_2(x))$ at $x_p$, using equations from the previous paragraph.

\subsubsection{MSAC lifetimes from the Breit-Wigner formula}
\label{subsubsec:BW}

In an alternative, quite different method, we also obtain the MSAC lifetimes from the Breit-Wigner formula (BW)~\cite{sakurai}. 
In the asymptotic regions, time-independent real-valued solutions on the classically allowed potentials are locally  of the form $\psi(x) = a \cos(k(x) x + \phi(W))$, with an energy-dependent phase $\phi$.
The asymptotic solution is a superposition of incident and back-scattered waves of respective forms $\exp(-i k x)$ and
$\exp(i (k x + 2 \delta)$, with the usual scattering phase shift $\delta$~\cite{sakurai}. It is thus seen that the phase $\phi$ in the time-independent solution equals the scattering phase, $\delta = \phi$. According to the BW formula, the decay rate of a MSAC, at the center of the scattering resonance,  is given by $\Gamma_{BW} = 2 (dW/d\phi) $, where the derivative is taken at a fixed location $x_B$ well outside the classically allowed range of the bound component of the MSAC wave function. Here we pick a location close to $x_{max}$; the exact value of $x_B$ is not important. The phase is then obtained from the classically-allowed tails of the wave functions $\psi_1(x)$ or $\psi_d(x)$ at $x_B$, in the diabatic and adiabatic representations, respectively, using
\begin{equation}
\phi(W)= \tan^{-1} \big(- \frac{1}{\psi(x_B)}\frac{d \psi(x)}{dx} \big|_{x=x_B} \big) + m \pi \quad ,
\end{equation}
where the integer $m$ is continually adjusted as a function of $W$ for continuity of $\phi(W)$. We subtract a background phase $\phi_0(W)$ that arises from the phase shift of the non-resonant solutions away from the MSACs and that is computed from
\begin{equation}
\phi_0(W)= \int_0^{x_B} \sqrt{2 (W - V_*(x))} dx \quad,
\end{equation}
where the potential $V_*(x) = - x/2$ in the diabatic and $V_*(x) = V_d(x)$ in the adiabatic representation. Note that for vanishing coupling, $V=0$, the phase would be that of an Airy-function solution~\cite{griffithsqm}).
The BW decay rates and lifetimes then become

\begin{equation}
\tau_{*,BW} = \frac{1}{\Gamma_{*,BW}} = 2 \frac{d (\phi_* - \phi_{*,0})} {dW} 
\quad ,
\end{equation}
where $* = nad $ and $* = ad$ for the diabatic and adiabatic representation, respectively. 

In summary of this subsection, we obtain four values for the nonadiabatic decay times of MSACs from solutions of time-independent two-component Schr\"odinger equations in diabatic and adiabatic representation, namely $\tau_{nad, FC}$, $\tau_{nad, BW}$, $\tau_{ad, FC}$, and $\tau_{ad, BW}$. As expected and shown below, these generally agree very well with each other, with the values from the adiabatic picture being more accurate due to the vanishing of the $A$- and $B$-coupling terms at large $|x|$.

\subsection{Time-dependent methods}
\label{subsec:tdep}

In our time-dependent computations, we utilize the scaled 
Hamiltonians in Eqs.~\ref{eq:hs} and~\ref{eq:ha} from Sec.~\ref{subsec:tindep} to find the MSAC lifetimes by propagating 
MSAC wave functions. For instance, in the adiabatic representation the time-dependent Schr\"odinger equation (TDSE) reads
\begin{eqnarray}
i \frac{\partial \psi_d (x, t)}{\partial t} & = & 
-\frac{1}{2} \frac{\partial^2 \psi_d (x, t)}{\partial x^2} + \tilde{V}_d(x) \psi_d (x, t) \nonumber\\ 
~ & ~ & + \big[B_{du}(x) + A_{du}(x) \frac{\partial}{\partial x} \big] \psi_u(x,t) \nonumber\\ 
i \frac{\partial \psi_u (x, t)}{\partial t} & = & 
-\frac{1}{2} \frac{\partial^2 \psi_u (x, t)}{\partial x^2} + \tilde{V}_u(x) \psi_u (x, t) \nonumber\\ 
~ & ~ & + \big[ B_{ud}(x) - A_{du}(x) \frac{\partial}{\partial x} \big] \psi_d(x,t) \quad.
\label{eq:BHSE1}
\end{eqnarray}
Here, we also have $B_{ud}(x) = - B_{du}(x)$ for all $x$. The TDSE in diabatic representation follows from
Eq.~\ref{eq:hs}.

As initial conditions for the MSAC wave functions at time $t=0$ in the diabatic and the adiabatic representations,  
we use the respective time-independent solutions obtained 
in Sec.~\ref{subsubsec:resonances}.
The MSAC wave functions from Sec.~\ref{subsubsec:resonances} exhibit oscillatory tails near the boundaries of the integration 
grid, as seen in Fig.~\ref{figure2}. To avoid numerical instability, the MSAC wave functions entered as initial states 
are set to zero between their outermost nodes and the respective spatial integration boundaries, $\pm x_{max}$. 

At the core of the TDSE method is to absorb the outgoing flux and to eliminate 
reflections from the boundaries~\cite{muga2004}. The wave-function norms then 
drop exponentially, thereby revealing the decay time of the MSAC entered as initial state.
The absorption is implemented by padding all diagonal potentials
with imaginary absorbing layers near the spatial integration boundaries at $\pm x_{max}$. 
The absorbing layers rise smoothly from zero at locations well-outside the classical turning points, $\pm x_l$, 
to a maximal value at $\pm x_{max}$. The utilized time-propagation method is a 
Crank-Nicolson scheme~\cite{cnaref} that is similar to schemes used in our recent work on tractor atom interferometry~\cite{tai} and Rydberg-ion molecules~\cite{duspayev2021nad}, where nonadiabatic transitions were quantitatively described. More details on the method can be found there. 
In the present work, the time-dependent computations are performed with a spatial-grid step size of $\Delta x = 10^{-3}$, the same as in the time-independent methods described in Sec.~\ref{subsec:tindep}, and a time-step size of $\Delta t = 10^{-3}$ (all in scaled units). 
We have checked that a reduction of $\Delta t$ does not significantly affect the lifetimes found for the MSAC wave functions.
The TDSE computations in the diabatic and adiabatic 
representations yield MSAC lifetimes denoted $\tau_{nad, TDSE}$ and $\tau_{ad, TDSE}$, respectively.

\section{Results}
\label{sec:res}

\begin{figure}[t!]
 \centering
  \includegraphics[width=0.48\textwidth]{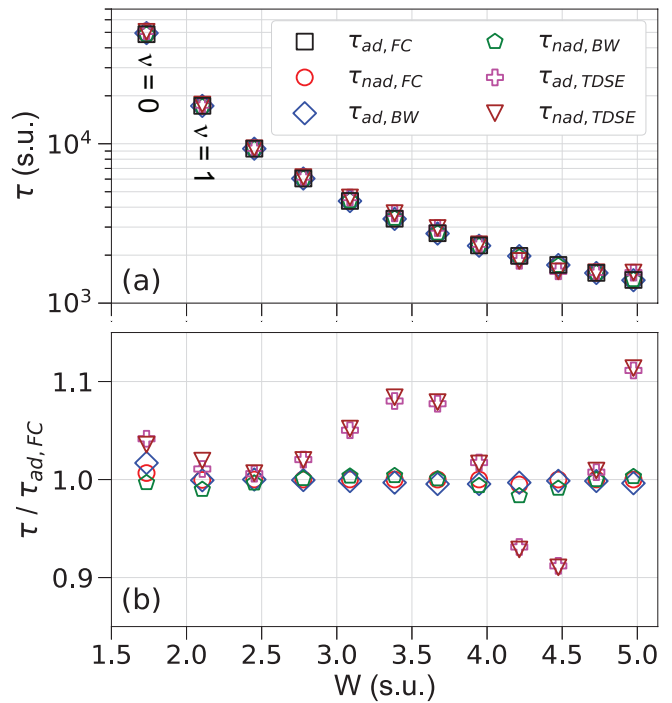}
  \caption{(Color online) MSAC lifetimes obtained with all six quantum methods, for vibrational states $\nu=0$ to 11, for an intermediate case of the coupling strength ($V=1.5275$~s.u.), plotted vs MSAC level energy $W$. The top panel demonstrates the overall agreement between all data and the drop-off of the lifetimes as a function of $\nu$. For improved visualization of small deviations, in the lower panel we show the ratios between the lifetimes from five quantum methods relative to $\tau_{ad, FC}$.  
  }
  \label{figure3}
\end{figure}

\subsection{Comparison of methods}
\label{subsec:rmethods}

In Fig.~\ref{figure3}~(a), we first present a comparison of results for a moderately adiabatic case, $V=1.528$, for $\nu=0$ to 11. The log-scale plot shows excellent 
agreement of lifetime data from all six methods over the entire range of $\nu$, over which the lifetime drops by about a factor of 30.
Among the methods, we consider the adiabatic wave-function flux results, $\tau_{ad,FC}$, to be the most accurate and 
precise for the following reasons.
The adiabatic analysis is less prone to numerical inaccuracy in the classically-forbidden tails of the wave functions, because the adiabatic couplings $A$ and $B$ drop off rapidly with increasing $|x|$ (see Figs.~\ref{figure1} and~\ref{figure2}, and arguments presented in Sec.~\ref{subsec:tindep}). This reduces the amplitude of the classically-forbidden tails, thereby alleviating their trend towards exponential divergence. {\afix{}Further, the} flux method
is insensitive to background-phase effects, which affects the BW method at low $V$ (see Sec.~\ref{subsec:rsurvey}). 

To exhibit small deviations of the results of the other five methods from $\tau_{ad,FC}$, in Fig.~\ref{figure3}~(b) we show the ratios $\tau_{*}/\tau_{ad,FC}$, with $*$ denoting the other methods. Importantly, the values for $\tau$ deviate by no more than 11$\%$ from $\tau_{ad,FC}$. The four results from the TIDSE agree to within 2$\%$ from each other, for $V=1.528$, with small deviations attributed to numerical inaccuracy and to the systematic inaccuracy of the BW method at low $V$ (see Sec.~\ref{subsec:rsurvey}). 
The computations based on the TDSE deviate by up to $11\%$ from $\tau_{ad,FC}$. This may be due to the
susceptibility of the time-dependent computations to imperfections of the absorbing-wall implementation, 
such as less-than-perfect absorption of the outgoing flux and spurious reflections.  
Indeed, for $\nu \leq 4$, where the absorbing walls are the farthest away from the high-amplitude regions of the MSAC wave functions, the TDSE lifetime results deviate by less than about 5$\%$ from the TIDSE results. It is also noted that the diabatic and adiabatic TDSE calculations differ by
less than about 1$\%$ from each other for all $\nu$-values. This indicates that numerical issues,
such as spatial-step or time-step sizes, introduce about the same, $\%$-level of uncertainty in the TDSE and the TIDSE calculations. 

Overall, the close agreement across the six methods in Fig.~\ref{figure3} proves the fundamental validity of all methods used. The quite good agreement between the TDSE and TIDSE calculations provides a particularly high level of validation, as the methods of how to extract the lifetimes from the TDISE and TDSE computations are quite different, yet both approaches yield very similar results.

\subsection{Lifetimes vs adiabaticity}
\label{subsec:rsurvey}

A main outcome of the work are the MSAC lifetimes over a wide range of the adiabaticity $V$ and the vibrational quantum number $\nu$. We have performed computations for a set of $V$-values ranging from $V=0.306$ (least adiabatic) to $V=2.75$ (most adiabatic). To assist with the interpretation of various regimes, we define the quality factor, $Q$, of the resonances as the angular frequency in harmonic approximation times the state's norm divided by the time derivative of the norm, 
or $ Q = \tau \omega = \tau / (2 \sqrt{V})$. There, $\omega$ is evaluated from the adiabatic potential $V_u(x)$ near $x=0$. Note $Q$ is unit-less and the same in scaled and physical units.

\begin{figure}[htb]
 \centering
  \includegraphics[width=0.48\textwidth]{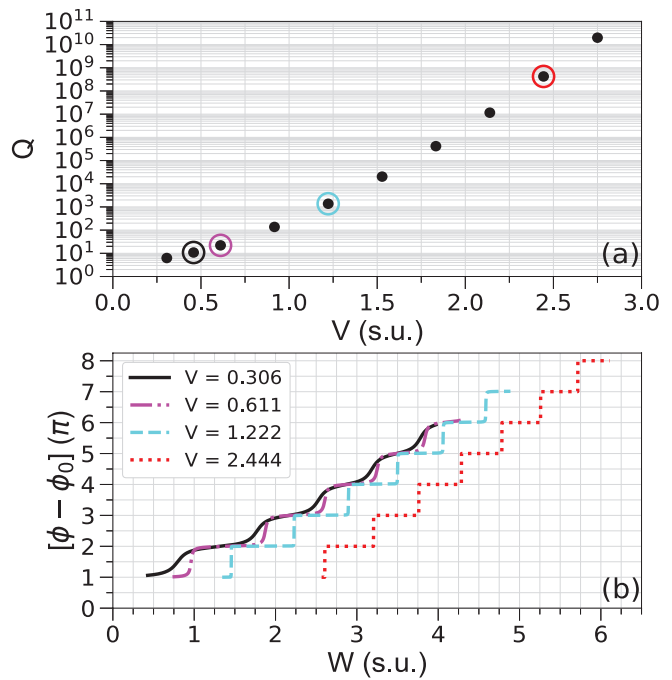}
  \caption{(Color online) (a) Quality factors, $Q$, as defined in the text, for the MSAC ground states, $\nu = 0$, vs coupling strength $V$. Circles around data points indicate the cases selected for panel (b).
  (b) Wave-function phases, $\phi(W)-\phi_0(W)$, vs energy for four values of $V$ selected in panel (a).}
  \label{figure4}
\end{figure}

In Fig.~\ref{figure4}~(a) we show the $Q$-values for the MSAC vibrational ground state, $\nu=0$, versus $V$. Noting that the number of oscillations after which the survival probability drops below 50$\%$ is approximately $Q/9$, it is seen that the ground-state MSACs may be considered only barely oscillatory for $0.3 < V \lesssim 0.6$, as in these cases it only takes a few oscillation periods or less for half of the ground-state MSAC population to decay. For $V \approx 1$ it already takes a few tens of oscillation periods before the ground-state MSACs are half decayed. However, as $V$ rises above about a value of 2, the ground-state MSACs quickly become highly stable against nonadiabatic decay. At the largest $V$-value tested, $V=2.75$, it takes $>10^9$ oscillation periods for half of the ground-state MSAC population to nonadiabatically decay [see Fig.~\ref{figure4}~(a)]. The rapid stabilization of MSACs as a function of $V$ is related to the factor of $-V^2$ in the exponential expression for the Landau-Zener tunneling probability (see Sec.~\ref{subsec:LZ}). 

The wide range of MSAC level damping is further visualized in Fig.~\ref{figure4}~(b), where we show four examples of 
the wave-function scattering phases, $\phi(W)-\phi_0(W)$, that are used for the calculation of BW lifetimes according to Sec.~\ref{subsubsec:BW}. At the MSAC energies, $W_\nu$, the phases exhibit rises in steps of $\pi$. The energy widths 
of the rises drop from a large fraction of the level spacing at $V=0.306$ to 
too narrow to be visible at $V=2.444$. Fig.~\ref{figure4}~(b) reiterates the vast range of nonadiabatic damping behavior that is seen over the range $0.3 < V < 2.444$, a range over which $V$ varies by about one and $Q$ by about ten orders of magnitude. 

Fig.~\ref{figure4}~(b) also shows that at the lowest $V$-values the resonances are wide enough and the slopes at the resonances, $d(\phi-\phi_0)/dW$, are small enough that background trends and cross talk between neighboring MSACs will affect the $d(\phi-\phi_0)/dW$-readings at the resonance centers, $W_{\nu}$. This makes lifetimes from the BW formula inaccurate at low $V$, as seen below.
Lifetimes from flux calculations are not susceptible to this type of inaccuracy. 

\begin{figure}[t!]
 \centering
  \includegraphics[width=0.48\textwidth]{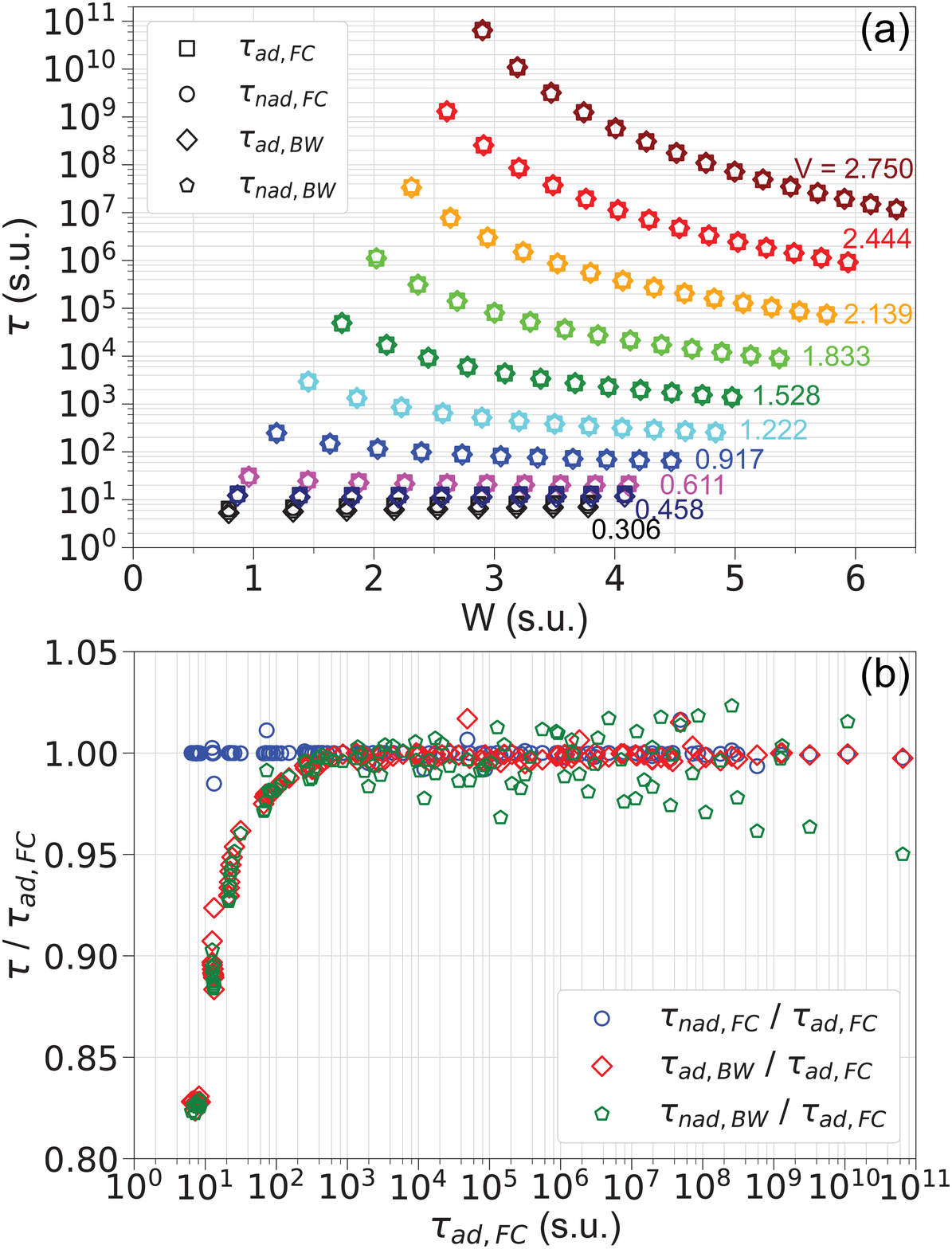}
  \caption{(Color online) (a) Results of lifetime calculations using the methods indicated in the legend, for a range of values of the coupling strength $V$. The calculations cover all MSAC resonances that are less than about 3.8 scaled energy units above the adiabatic-potential minimum of $V_u$ at $x=0$. (b) Ratios between the lifetimes shown in (a) and $\tau_{ad,FC}$ vs $\tau_{ad,FC}$. The plot includes data points for
  all $V$- and $\nu$-values also shown in (a).}
  \label{figure5}
\end{figure}

In Fig.~\ref{figure5} we show lifetime results from TIDSE computations for ten values of $V$ for MSACs within an energy range of about 3.8~s.u. from the potential minima of $V_u$. (The computationally more intensive TDSE 
computations were performed only for the intermediate case of $V=1.528$.) 
Fig.~\ref{figure5}~(a) demonstrates good agreement between the TIDSE methods over a wide range of conditions. 
For all $\nu$-values studied, the MSAC lifetimes increase by six to ten orders of magnitude, as $V$ is increased from 0.306 to 2.75. 
In the following, we discuss the dependence of the lifetimes on $\nu$ in several regimes of $V$. 

In the nonadiabatic regime, $V \lesssim 0.6$, the lifetime barely depends on the vibrational quantum number, $\nu$, and for the least-adiabatic case, $V = 0.306$, the lifetime actually increases with $\nu$. This behavior, which may seem counter-intuitive at first, reflects the fact that for $V \ll 1$ the anti-crossing half width, $x_w=2 V$, which is an estimate for the reach of the $A$- and $B$-couplings, 
only is a fraction of the spatial extent of the MSAC wave function on $\tilde{V}_u$, as seen above in 
Fig.~\ref{figure1}~(a). As a result, for $V \ll 1$ the spatial extent of the interaction range that causes the nonadiabatic decay, measured relative to the wave-function extent, decreases with increasing $\nu$, leading to an increase in lifetime with increasing $\nu$. This mechanism becomes more transparent in an analysis based on Fermi's Golden Rule (see Sec.~\ref{subsec:FGR}). Arguing semi-classically, one may say that at the lowest $V$-values studied the lifetime increases with $\nu$ because with increasing $\nu$ the MSAC particle spends less of its time in the anti-crossing region. It is noted that increasing $\nu$ for the purpose of increasing the MSAC lifetime is not a useful concept to generate long-lived 
MSACs (for atom trapping, for instance), because of the generally very low $Q$-values at $V \lesssim 0.6$ [see Fig.~\ref{figure4}~(a)]. 

For $V \gtrsim 1.2$, the MSAC resonances become increasingly adiabatic, with $Q$-values beginning to range above 100.
In the adiabatic regime, the MSAC lifetimes decrease with increasing $\nu$, which is opposite to the trend that is seen in the nonadiabatic regime.  The decrease of $\tau$ with increasing $\nu$ accelerates with increasing $V$; at $V=2.75$, the largest value studied, the lifetime ratio between $\nu=0$ and $\nu=10$ exceeds a factor of 1000.
In order to understand this behavior, one may first compare the relative importance of the $A$- and $B$-coupling terms in the adiabatic representation. It is found in Sec.~\ref{subsec:FGR} that the $A$-term is quite dominant. As a consequence, at sufficiently large $V$, the gradient of the trapped wave function, $\frac{\partial} {\partial x} \psi_u$, averaged over the wave-function extent, factors decisively into the nonadiabatic coupling strength. This means that, at the larger $V$-values, 
the lifetime should drop with increasing $\nu$, as observed. Noting that wave-function gradient and classical velocity are related,  the velocity dependence of the Landau-Zener equation predicts the same trend (see Sec.~\ref{subsec:LZ}). 

Next, we discuss the deviations between the lifetimes obtained with the TDISE methods.
For visibility of small deviations, we display the ratios $\tau_{*}/\tau_{ad, FC}$ on a fine scale in Fig.~\ref{figure5}~(b). The adiabatic and nonadiabatic flux-calculation results agree very well in all regimes.
We reiterate that $\tau_{ad, FC}$ is still considered to be the most accurate and precise (see Sec.~\ref{subsec:rmethods}). For $\tau_{ad,FC} \gtrsim 200$, which roughly corresponds with $V \lesssim 1$, the BW data also agree well. However, for $\tau_{ad,FC} \lesssim 200$ they yield up to about 20$\%$ shorter lifetimes than the flux methods. It is also noted that the two BW results from the diabatic and adiabatic representations still 
agree very well with each other. The systematic deviation of the BW from the flux-calculation data at low $V$ (nonadiabatic regime)
may be attributed to the facts that at low $V$  
neighboring BW resonances begin to cross-talk, and that background phase slopes become a significant fraction of the slopes $d (\phi - \phi_0)/dW$ at the resonance centers [see Fig.~\ref{figure4}~(b)], rendering the BW data less accurate at low $V$. It is further seen that the numerical noise of the diabatic BW calculations can reach $5\%$ at large $V$, where the resonances become extremely narrow and the computation of the slopes $d (\phi - \phi_0)/dW$ becomes less accurate. Notwithstanding, the overall good agreement, seen on the fine scale in Fig.~\ref{figure5}~(b),
validates methods and results across the entire $V$- and $\nu$-regimes studied.

\section{Approximation methods}
\label{sec:approx}

\subsection{Perturbation theory}
\label{subsec:FGR}

The adiabatic representation lends itself to a perturbative description of nonadiabatic decay~\cite{Burrows_2017, mandal2018}. In this approach, we find bound MSACs on $\tilde{V}_u(x)$ used in Eq.~\ref{eq:ha}, neglecting the nonadiabatic $A$- and $B$-couplings (but keeping the diagonal $B$-terms). 
These states differ from the true MSACs in that they are infinitely-long-lived,
and in that their energies, $W_{\nu, FGR}$, are up to 0.07 scaled units below the true resonance energies, $W_{\nu}$.
The energy deviations are most notable at small coupling $V$, where the off-diagonal non-adiabatic terms are large and cause the largest shifts 
$W_{\nu}- W_{\nu, FGR} $. We denote the wave functions of the coupling- and decay-free approximations of the MSACs as $\psi_{u, \nu, FGR} (x)$. The $\psi_{u, \nu, FGR} (x)$ are weakly coupled to the continuum of free-particle states on the potential $\tilde{V}_d(x)$. The solutions on $\tilde{V}_d(x)$ are, asymptotically, identical with Airy functions~\cite{griffithsqm}. Factoring in that on $\tilde{V}_d(x)$ the wave functions extend to both $\pm \infty$, as opposed to just one side on a linear potential, we normalize the free states such that the amplitude of their oscillatory tails at large positive $x$ is 
\begin{equation} 
\psi_{d, W, FGR} (x) \approx  \sqrt{ \frac{1}{\pi}} \Big| x + 2 W \Big|^{-1/4}  \quad .
\end{equation}
There, $W$ is the level energy. The solutions $\psi_{d, W, FGR} (x)$ normalized in that way are
orthonormal in unit energy, {\sl{i.e.}} it is
$\langle \psi_{d, W', FGR} \vert \psi_{d, W, FGR} \rangle = \delta(W-W')$. According to Fermi's Golden Rule (FGR), the transition rate from
$\vert \psi_{u, \nu, FGR} \rangle $ to $\vert \psi_{d, W, FGR} \rangle $ then is given by $\Gamma_{FGR} = 2 \pi |M|^2$, with a matrix element 
\begin{eqnarray} 
M & =  & \langle \psi_{d, W, FGR} \vert \hat{B}_{du} +  {\rm i} \hat{A}_{du} \hat{p}_x \vert \psi_{u, \nu, FGR} \rangle \nonumber \\
& = & \int \psi^*_{d} (x) \Big[ B_{du}(x) + A_{du}(x) \frac{d}{dx}  \Big] \psi_{u}(x) dx \quad ,
\label{eq:FGR}
\end{eqnarray}
where we abbreviate $\psi_u(x) = \langle x \vert \psi_{u, \nu, FGR}\rangle$ and $\psi_d(x) = \langle x \vert \psi_{d, W, FGR} \rangle$. The free-particle energy in the integral equals that of the quasi-bound state, $W = W_{\nu, FGR}$. Also, here all $\psi(x)$ are real, and the integration range is limited by the range of $\psi_u(x)$. Since in the present problem $B_{du}(x)$ and $d/dx$ have odd and $A_{du}(x)$ has even parity in $x$, even $\psi_{u} (x)$ decay into odd solutions $\psi_{d} (x)$ and vice versa. The FGR lifetimes then are 
\begin{equation} 
\tau_{FGR} = 1 / ( 2 \pi |M|^2 ) \quad .
\label{eq:FGR2}
\end{equation}

\begin{figure}[t!]
 \centering
  \includegraphics[width=0.48\textwidth]{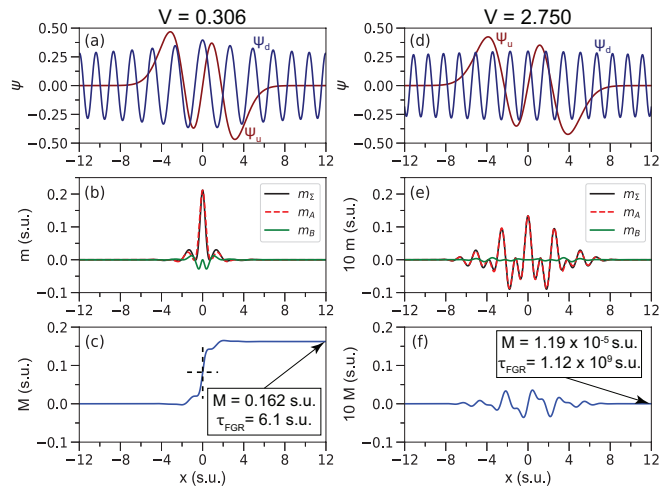}
  \caption{(a, d) Wave functions $\psi_u(x)$ and $\psi_d(x)$ for MSAC $\nu=3$ at $V=0.306$ and $V=2.75$, respectively. 
  (b, e) Transition amplitudes $m_A(x)$, $m_B(x)$ and $m_{\Sigma}(x)$, defined in the text, for the states in (a, d), respectively. 
  (c, f)  $M(x) = \int^{x} m_{\Sigma}(x') dx'$ for the states in (a, d), respectively. The transition matrix element $M$ in Fermi's Golden Rule, given by $M(x)$ at the right margins, and the FGR lifetimes for the state, $\tau_{FGR}$, are indicated in the boxes.}
  \label{figure6}
\end{figure}

The FGR calculation is visualized in Fig.~\ref{figure6} for a small and a large $V$-value, for the case  $\nu=3$. While the bound and free wave functions, $\psi_u(x) = \langle x \vert \psi_{u, \nu=3, FGR}\rangle$ and $\psi_d(x) = \langle x \vert \psi_{d, W, FGR} \rangle$, look quite similar in the two cases [see Fig.~\ref{figure6}~(a,d)], the coupling matrix elements are very different. We define the matrix-element ``densities''
$m_A(x) = \psi_{d} (x) A_{du}(x) \frac{d}{dx} \psi_{u}(x)$, 
$m_B(x) = \psi_{d} (x) B_{du}(x) \psi_{u}(x)$, and the coherent sum $m_{\Sigma}(x)=m_A(x) + m_B(x)$, and display these functions in Fig.~\ref{figure6}~(b,e). For $V=0.306$ the $m$-densities are large, localized to within just the central lobe of $\psi_u(x)$, and largely uni-polar, whereas for $V=2.75$ the $m$-densities are weak, spread-out over the entire reach of $\psi_u(x)$, highly oscillatory and bi-polar. In both cases, $m_A(x)$ is much larger than $m_B(x)$ in magnitude, on average. As a result, for small $V$ the nonadiabatic decay is fast and largely driven by couplings localized to within a small, interior region of $\psi_u(x)$, whereas for large $V$ the nonadiabatic decay is slow and spread-out over the 
entire range of $\psi_u(x)$. These observations validate statements that we have made in Sec.~\ref{subsec:rsurvey} with regard to the $\nu$-dependence of the MSAC lifetimes. 

In Fig.~\ref{figure6}~(c,f) we show the integrals of $m(x)$, whose asymptotic values, $M=M(x_{max})$, give the FGR lifetimes according to Eq.~\ref{eq:FGR2}. Due to symmetry, the integral $M(x)$ has odd parity about a symmetry point at $x=0$ [crosshair in Fig.~\ref{figure6}~(c)], 
and it is $M = M(x_{max}) = 2 M(x=0)$. For low and moderate values of $V$, the large amplitudes and the somewhat uni-polar characteristics of $m(x)$ lead to numerically stable results for $M$ and $\tau_{FGR}$. At large $V$, however, the integral in Eq.~\ref{eq:FGR} is numerically challenging because of the bipolar and highly oscillatory behavior of $m_{\Sigma}(x)$. It is seen in Fig.~\ref{figure6}~(f) that at large $V$ the integral $M = \int m_{\Sigma}(x) dx$ comes down to a very small, nearly-vanishing remainder after integration, as evidenced by the fact that $M(x)$ has a near-perfect zero crossing at $x=0$, leading to a very small matrix element $M$. To get converging values for $M$, at the largest $V$-values studied we had to decrease the spatial step size in the wave-function computations and in the integral for $M$ by a factor of up to about 100 relative to the step size used 
in the non-perturbative methods. Nevertheless, even at the largest $V$ considered the FGR computations are still quite fast because the wave functions to be computed are scalar.   

\begin{figure}[t!]
 \centering
  \includegraphics[width=0.48\textwidth]{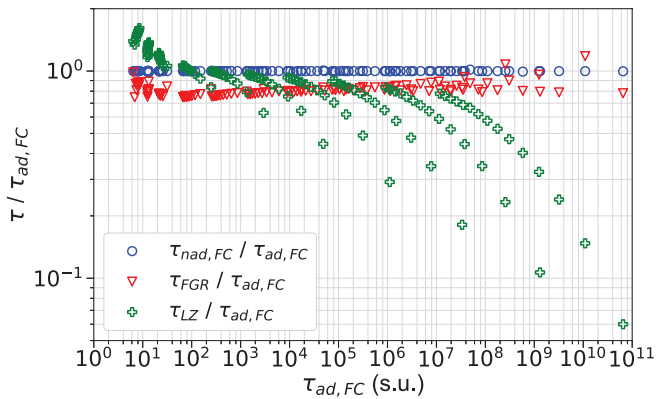}
  \caption{Ratios between lifetimes from a perturbative calculation based on Fermi's Golden Rule, $\tau_{FGR}$, and from the semi-classical Landau-Zener approximation, $\tau_{LZ}$ (see Sec.~\ref{subsec:LZ}), divided by the lifetime $\tau_{ad,FC}$ from the exact calculation, versus $\tau_{ad,FC}$. For reference, we also show
  $\tau_{nad,FC}/\tau_{ad,FC}$. 
  The data points encompass all MSACs also shown in Fig.~5.} 
  \label{figure7}
\end{figure}

In Fig.~\ref{figure7} we present the ratio $\tau_{FGR}/\tau_{ad,FC}$ for all values of $V$ and $\nu$ also shown in Fig.~\ref{figure5}. 
As in Figs.~\ref{figure3}~(b) and~\ref{figure5}~(b), $\tau_{ad, FC}$ is used as a reference because 
the non-perturbative adiabatic wave-function flux calculation is the most accurate and precise. 
The lifetime ratios are plotted on a log scale covering about two decades, which is fine enough to 
observe relative deviations as small as about 1$\%$ and wide enough to also 
cover relative deviations for a Landau-Zener model (see Sec.~\ref{subsec:LZ}). 
The $\tau_{FGR}/\tau_{ad,FC}$-ratios, plotted in Fig.~\ref{figure7} versus $\tau_{ad,FC}$, 
follow a quite well-defined trend line at 0.1 to 0.3 below unity, 
with the lowest deviations occurring in the nonadiabatic and adiabatic limits on the left and 
right margins of the plot, respectively. At the largest $\tau_{ad,FC}$-values, corresponding to large $V$- and low $\nu$-values,  
there is additional numerical noise on the order of $\pm 0.1$, caused by the delicate 
nature of the $M$-matrix elements at large $V$ (see Fig.~\ref{figure6} and related discussion).     

The FGR approach in this work differs from typical applications of FGR in which the
wave functions are perturbation-independent and the perturbation has a tunable strength. 
In contrast, in the present case the perturbation $V$ is fixed for a given set of
wave functions $\psi_u(x)$ and $\psi_d(x)$, and the wave functions themselves depend on the fixed perturbation $V$. The 
matrix-element ``densities'' $m(x)$ have a complex spatial structure and are considered in first order only. 
The deviations of the FGR from the non-perturbative results are notable, albeit not exceeding about $30\%$.
A practical concern relies in the fact that at large $V$ the spatial step size in the FGR calculation of the matrix element $M$ has to be set 
very small to achieve convergence, due to the delicate nature of the $M$-integral at large $V$ (see Fig.~\ref{figure6}).

\subsection{Landau-Zener model}
\label{subsec:LZ}

For a semi-classical estimate of MSAC lifetimes using the
Landau-Zener equation, we use a Landau-Zener tunneling ``attempt rate'' of twice the vibrational frequency, which gives an attempt rate of
$R(\nu) = (W_{\nu+1} - W_{\nu-1})/\pi$ (scaled units). The LZ coupling equals $V$ and the differential slope of the diabatic potentials equals $s=1$, in scaled units. For a fixed particle velocity, $v$, the Landau-Zener tunneling probability is
$P_{LZ} = \exp(-2\pi V^{2}/( s \, v))$, and the lifetime $\tau_{LZ} = 1/(R P_{LZ})$, in scaled units. 
Assuming that a semi-classical picture with a point-particle velocity $v$ suffices 
to describe the quantum problem of interest, one needs a rule for how to get $v$.
From Fourier transforms of MSAC wave functions in any representation (diabatic or adiabatic),  
one expects and finds that $v$ could be on the order of $\sqrt{W_\nu-V}$, which 
also accords with the classical virial theorem for a harmonic oscillator. 
Further, classically the velocity peaks at $v = \sqrt{2(W_\nu-V)}$ at the crossing. 
For the largest $V$ and lowest $\nu$ studied in this work, 
these $v$-values produce $\tau_{LZ}$-values that are about 20 orders of magnitude too long. 
As the exponent in the Landau-Zener tunneling probability is $\propto -1/v$,   
we may surmise that the high-velocity wings in the Fourier transforms of the MSAC wave functions 
govern the LZ decay rate. Empirically, one finds that $v = \sqrt{2 W_\nu}$, used in the following, overall leads to the best 
LZ estimates for the MSAC lifetimes (that can still be several orders of magnitude off).

The deviations of $\tau_{LZ}$ from quantum calculations are shown in Fig.~\ref{figure7} in terms of $\tau_{LZ}/\tau_{ad,FC}$. 
It is seen that, over our range in $V$ and $\nu$ studied, the LZ model may serve as a very rough guideline to predict MSAC lifetimes, as the 
$\tau_{LZ}$-values stay within a factor of about 
20 from $\tau_{ad,FC}$. 
The inaccuracy of the $\tau_{LZ}$-values accelerates in the adiabatic region (large $\tau_{ad,FC})$. The strong $\nu$-dependence of $\tau_{LZ}/\tau_{ad,FC}$, seen especially in the adiabatic region, reiterates that we have
no well-founded rule for the classical velocity $v$. As such,
the poor overall agreement of $\tau_{LZ}$ with the quantum results reflects the 
fact that a semi-classical model applied on a problem in the quantum domain cannot be expected to be accurate. 

Considering quantum-classical correspondence, we add that with increasing $\nu$ our model system becomes more classical, and with 
decreasing $V$ the nonadiabatic transitions become relatively well-localized in the spatial region near $x = 0$. 
As a result, for $V \lesssim 1$, and for $V \gtrsim 1 $ and $\nu$ exceeding a $V$-dependent 
limit evident from Fig.~\ref{figure7}, the $\tau_{LZ}$-values deviate by less than about 50\% from the corresponding $\tau_{ad,FC}$-values, and the agreement improves with increasing $\nu$. These observations accord with the expectation that quantum-classical 
correspondence should occur in the limit of large quantum numbers.

\section{Conclusion}
\label{sec:concl}

We have computed nonadiabatic lifetimes of metastable states on symmetric avoided crossings. Among six non-perturbative quantum methods, the results of which generally agree well, a wave-function flux method implemented in the adiabatic representation is the most accurate and precise, with lifetime uncertainties estimated at about $1\%$. Using the given relations between scaled and physical units, the results are portable to a variety of applications, including Rydberg molecules~\cite{shafferreview, feyreview} and atom trapping and guiding on dressed potentials~\cite{Garraway_2016}. 

In addition to providing accurate, non-perturbative lifetime data, our comparisons have shown that time-dependent perturbation theory in first order, applied to states in the adiabatic representation, with the nonadiabatic coupling terms treated as a perturbation, 
yields approximate lifetimes that deviate by less than about 30$\%$ from the non-perturbative values. Semi-classical estimates based on the Landau-Zener tunneling formula were generally found to be quite inaccurate. This is especially the case for vibrational ground states in the adiabatic (long-lifetime) regime, which are states of paramount relevance in atom trapping and guiding. 
Expanding on earlier works in atom trapping~\cite{Burrows_2017} and 
Rydberg molecules~\cite{duspayev2021nad}, the non-perturbative methods tested in the present work
can be generalized to problems with more than two adiabatic potentials with non-linear spatial dependence and variable mutual couplings.

\section*{ACKNOWLEDGMENTS}
The work was supported by the NSF Grant No. PHY-2110049 and in part through computational resources and services provided by Advanced Research Computing at the University of Michigan, Ann Arbor. 

\bibliography{references}
\end{document}